%% file: paper.tex
\documentclass{article}

    \PassOptionsToPackage{numbers, compress}{natbib}

\usepackage[preprint]{neurips_2026}

\usepackage[utf8]{inputenc}
\usepackage[T1]{fontenc}
\usepackage[table, dvipsnames]{xcolor}
\usepackage[
    colorlinks=true,
    linkcolor=red,    
    citecolor=blue,   
    urlcolor=magenta  
]{hyperref}
\usepackage{bookmark}
\usepackage{url}
\usepackage{booktabs}
\usepackage{amsfonts}
\usepackage{amssymb}
\usepackage{nicefrac}
\usepackage{microtype}
\definecolor{lightblue}{rgb}{0.82, 0.88, 1.0} 
\usepackage{wrapfig}
\usepackage{amsmath}
\usepackage{amsthm}
\usepackage{graphicx}
\usepackage{array}
\usepackage{tabularx}

\title{Adversarial Prompts for Acceptance Collapse in Speculative Decoding}

\author{
 \textbf{Run Wang\textsuperscript{1}},
 \textbf{Chaoyi Zhou\textsuperscript{1}},
 \textbf{Xi Liu\textsuperscript{1}},
 \textbf{Yi Zhu\textsuperscript{2}},
 \textbf{Amir Salarpour\textsuperscript{1}},
\\
 \textbf{Pedram MohajerAnsari\textsuperscript{1}},
 \textbf{Zhi-Qi Cheng\textsuperscript{3}},
 \textbf{Feng Luo\textsuperscript{1}},
 \textbf{Siyu Huang\textsuperscript{1}}
 \textbf{Mert D. Pesé\textsuperscript{1}$^\ast$}
\\
 \textsuperscript{1}Clemson University
\\
 \textsuperscript{2}Wayne State University
\\
 \textsuperscript{3}University of Washington
\\
}

\begin{document}

\maketitle{
\renewcommand{\thefootnote}{}
\footnotemark
\footnotetext{$^\ast$ Corresponding author.}
}

\input{Sections/Abstract}
\input{Sections/Intro}
\input{Sections/Background}

\input{Sections/Threat}
\input{Sections/Methodology}
\input{Sections/Eval}

\input{Sections/Discussion}
\input{Sections/Conclusion}

\bibliographystyle{unsrt}
\bibliography{references}
\input{Sections/Appendix}

\newpage

\end{document}

%% file: Sections/Abstract.tex
\begin{abstract}
Lossless acceleration schemes, such as speculative decoding, promise significant inference speedups by relying on dynamic token-level alignment between a draft and a target model. However, this guarantee of semantic equivalence masks a severe operational vulnerability: draft-target alignment can be systematically attacked. In this paper, we introduce ADSD, which, to the best of our knowledge, is the first prompt-suffix attack that collapses verifier acceptance by pushing draft probability mass toward tokens the target is unlikely to accept. ADSD uses Soft-Collapse, a verifier-aligned surrogate derived from the asymmetric speculative acceptance rule, together with a target-preservation objective that discourages obvious task corruption. ADSD successfully generates highly effective adversarial suffixes. On the GSM8K dataset, our attack increases the mean sample time by 62.3\% while preserving the task quality. We further show that this vulnerability exists across different domains, speculative decoding strategies, and model architectures.
\end{abstract}

%% file: Sections/Intro.tex
\section{Introduction}

Inference efficiency has become a significant systems bottleneck for large language models (LLMs)~\citep{achiam2023gpt, touvron2023llama, bai2023qwen}. Currently, different methods have been proposed to increase inference efficiency while reducing computational cost, such as pruning~\citep{sun2023simple,frankle2018lottery} and quantization~\citep{xiao2023smoothquant,shen2020q}. However, these methods can introduce performance degradation. Consequently, lossless acceleration techniques, particularly speculative decoding~\citep{xia2023speculative, leviathan2023speculative}, have emerged as a highly promising and widely adopted design methodology. In this paradigm, a rapid drafting mechanism, whether an independent smaller model or a lightweight auxiliary head sharing the target's hidden states, proposes candidate tokens. A larger target model then verifies these proposals in parallel, allowing the system to gain substantial speedups without altering the target distribution and overall system performance~\citep{miao2024specinfer,cai2024medusa,huang2025adaspec,xia2024survey,zhou2026hsd,sun2024block,qin2025optimized,li2024eagle,li2024eagle2,li2025eagle3,huang2025litevlmlowlatencyvisionlanguagemodel,biran2026aws,kumar2026speculative,chen2023accelerating,wu2024snakes,liu2024sdsat}.

\begin{figure}[t]
\centering
\includegraphics[width=0.9\linewidth]{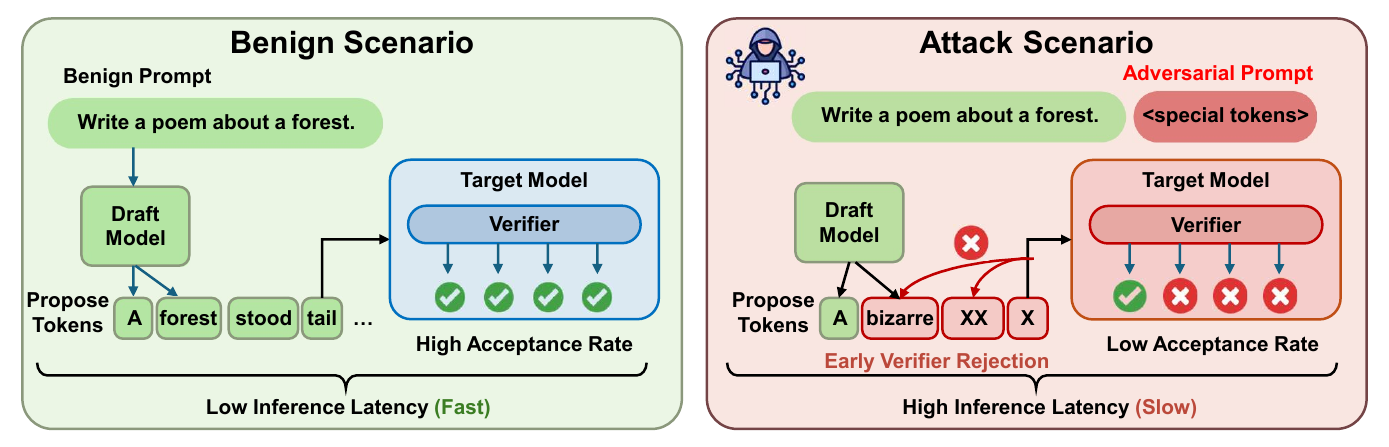}
\caption{Threat model overview. An attacker does not modify model
weights or decoding logic. Instead, the attacker inserts a short suffix that
collapses draft-target agreement and forces the speculative verifier to perform
substantially more work.}
\label{fig:threat-model}
\end{figure}

The efficiency gains of speculative decoding critically depend on high verifier acceptance. As illustrated in Figure~\ref{fig:threat-model}, when draft tokens are accepted, a single target-model
pass can commit multiple tokens; however, if an early draft token is rejected, the remaining draft tokens are discarded and have to be re-generated again by the draft model, resulting in another round of drafting and verification. Thus, a lower acceptance rate not only reduces the nominal acceleration, but also turns drafting and verification into repeated wasted computation, increasing end-to-end latency and serving cost. Most prior work on
speculative decoding focuses on improving acceptance under benign inputs, such as improving the draft-target alignment or redesigning the verification
procedure~\citep{liu2023online,goel2024direct,agrawal2024adaedl,bachmann2025judge,
wang2025alignment,yan2025decoding,zhou2026hsd}. However, this line of work largely
overlooks the security implications of the acceptance mechanism. Specifically, attackers can
intentionally manipulate verifier acceptance to increase the response latency and computational resource consumption, compromising the availability of LLM services.

To the best of our knowledge, this work is the first to study vulnerabilities in the internal acceptance mechanism of speculative decoding.
Recent security studies on speculative decoding mostly examine side-channel leakage and safety-aware decoding~\citep{wei2024spills,wang2025ssd}, while broader availability attacks against LLM typically aim to inflate latency or cost by elongating outputs or exploiting schedulers and guardrails outside the core decoding algorithm~\citep{shumailov2021sponge,zhang2024safeguard,dong2024engorgio,wang2026latencydos}. In contrast, the proposed ADSD targets the core mechanism in speculative decoding itself, i.e., the internal drafting-verification process. As shown in Figure~\ref{fig:threat-model}, ADSD constructs an adversarial suffix that induces the verifier to reject most draft tokens, forcing repeated drafting-verification rounds. The key challenge is to sufficiently degrade verifier acceptance to increase drafting-verification rounds, while preserving the LLM response quality so that the attack remains stealthy. To address this challenge, we propose a Soft-Collapse objective and a target-preservation objective, and a gradient-guided discrete prompt-search algorithm that jointly optimizes both objectives.

Experimental results show that ADSD introduces performance issues in speculative decoding. On a matched GSM8K benchmark, our attack suffix reduced speculative decoding efficiency by increasing mean sample time from 26.05s to 42.29s (+62.3\%), while preserving the task quality. Furthermore, we evaluate this attack design on different datasets and architectures, and show that the attack remains consistently effective in the evaluated settings.

\begin{itemize}
    \item To the best of our knowledge, we are the first to exploit prompt-induced verifier collapse as a speculative decoding risk, distinct from conventional safety failures or output-length inflation attacks.

    \item We develop ADSD, a prompt-suffix search method powered by the
    Soft-Collapse surrogate objective, which targets the asymmetric acceptance rule of
    speculative verification.

    \item We show that short suffixes can substantially reduce speculative
    decoding efficiency on GSM8K while preserving task accuracy, and that the effect persists across several decoding strategies, model families, and task
    domains.
\end{itemize}

%% file: Sections/Background.tex
\section{Background and Related Work}
\label{sec:background}

\paragraph{Speculative decoding.}
Speculative decoding accelerates autoregressive generation by pairing a rapid drafting mechanism with a target verifier. While initially formulated with independent draft models~\citep{leviathan2023speculative}, modern architectures increasingly employ auxiliary drafting heads (e.g., EAGLE~\citep{li2024eagle,li2024eagle2,li2025eagle3}, Medusa~\citep{cai2024medusa}) that reuse the target's hidden states to improve alignment without loading separate weights. In either approach, the drafter proposes candidate tokens that the target verifies in parallel; accepted tokens are committed, while rejected ones are replaced with target-sampled corrections. Because this foundational draft-then-verify pattern drives a wide array of speculative systems~\citep{miao2024specinfer,cai2024medusa,xia2024survey,yan2025decoding,zhou2026hsd,li2024amphistabidirectionalmultiheaddecoding}, inference efficiency depends directly on the distributional overlap between the draft and target models. High agreement yields long accepted draft blocks, whereas divergence forces frequent verifier interventions.

The speculative-decoding literature treats low acceptance as an
efficiency problem to be repaired. Alignment-centric approaches adapt or train
the draft model so that its token proposals more closely match the target
\citep{liu2023online,goel2024direct,wang2025alignment}. Verification-centric
approaches instead redesign the accept-reject mechanism itself, for example by
predicting likely failures and halting drafting adaptively, or by recovering
useful work even under partial disagreement
\citep{agrawal2024adaedl,bachmann2025judge,yan2025decoding,zhou2026hsd,liu2026speculativedecodingperformanceillusion}. These
methods differ substantially in mechanism, but they share the same premise:
prompts are treated as benign, and the system designer's goal is to preserve
acceptance under ordinary inputs. Our work begins exactly where that premise
breaks down: we study the prompt itself as the attack surface.

\vspace{-10pt}
\paragraph{Safety of speculative decoding.}
Speculative decoding has begun to receive security scrutiny, but
the existing papers target different failure modes. Recent work shows that
speculation traces can leak information through timing and packet-size side
channels~\citep{wei2024spills}. Another line of work uses speculative decoding
as a safety mechanism, where a smaller aligned model steers the larger one away
from harmful outputs while retaining some efficiency gains
\citep{wang2025ssd}. However, neither line studies a prompt-side adversary 
whose explicit goal is to collapse verifier acceptance and thereby degrade speculative-decoding acceleration. Recent empirical work also shows that speculative 
decoding is not always beneficial: for small language models or mismatched 
draft‑target pairs, acceptance can naturally degrade~\citep{mainardi2026empirical}.
\vspace{-10pt}
\paragraph{LLM inference-cost and availability attacks.}
Attacks that inflate the latency, energy, or compute cost of general LLMs have been explored. Sponge examples show
that carefully constructed inputs can force neural networks toward worst-case
energy-latency behavior~\citep{shumailov2021sponge}. In autoregressive LLMs,
recent attacks exploit guardrail mediation, output length, or scheduler
behavior to amplify per-request serving cost
\citep{zhang2024safeguard,dong2024engorgio,wang2026latencydos}. ADSD belongs
to this availability-oriented family, but its mechanism is substantially
different. Rather than elongating outputs or attacking the serving stack
around the model, it attacks the acceptance process inside speculative decoding itself.

%% file: Sections/Threat.tex
\section{Threat Model and Impact}
\label{sec:threat}

\subsection{Threat Model}
A key operational fragility of speculative decoding lies in its structural reliance on distributional alignment. While the system guarantees semantic equivalence to the target model, its performance is entirely bottlenecked by the drafting mechanism's ability to approximate the target's latent manifold. Under benign usage, this approximation holds. However, this creates a deeply coupled dynamical system where any engineered divergence between the draft distribution $q$ and target distribution $p$ immediately translates into wasted compute. The attacker's objective is to exploit this divergence, crafting inputs that systematically force the draft model into generating overconfident but consistently rejected drafted tokens, while making the output quality appear normal. We formalize this threat as a prompt-suffix optimization attack against the speculative decoding pipeline.

\noindent\textbf{Attacker Capabilities.} The attacker operates strictly in the user space, restricted to injecting a short, discrete text suffix $z$ of length $m$ into an otherwise benign prompt. The adversary cannot manipulate model weights, KV-cache states, temperature settings, batching policies, or any serving-stack schedulers. The attack relies entirely on exploiting token interactions within the transformer forward pass.

\noindent\textbf{Attacker Knowledge.} We assume a white-box setting where the attacker has offline access to the drafting distribution $q$ (whether generated by an independent model or a shared-feature head), the target model $p$, the shared tokenizer, and the specific speculative verification mechanism (e.g., standard tokenwise acceptance). This offline access is used to optimize the adversarial suffix prior to deployment. 

\noindent\textbf{Attack Objective.} The goal is to maximize the verifier's computational burden per request, functionally erasing the speculative speedup, while remaining stealthy. If the prompt simply triggers gibberish or a system refusal, the latency degradation is easily flagged by standard quality monitors and discarded. 

\subsection{Systemic Impact}
The impact of this vulnerability extends well beyond single-request latency, because modern LLM serving stacks are explicitly engineered around throughput, batching efficiency, and per-token cost control rather than semantic correctness alone. Systems such as vLLM and TGI achieve high utilization by combining memory-efficient serving with shared continuous batching, while speculative decoding is increasingly promoted in both research systems and production guidance as a practical way to reduce decode-side cost and latency~\citep{kwon2023pagedattention,huggingface2026tgi,reboul2025continuousbatching,biran2026aws}. The same economics are reflected in commercial platforms, where providers expose rate limits, provisioned throughput, and token-based pricing as first-class operational constraints~\citep{openai2026pricing,openai2026ratelimits,azure2025latency,google2025provisioned}. Under this deployment model, ADSD is not merely a slowdown attack on one prompt: by forcing persistent verifier rejection, it converts a lossless acceleration path into wasted target-model work, inflates the real cost of a request, and creates a prompt-side denial-of-wallet surface for any downstream application paying per token or operating under fixed inference quotas. Moreover, because batched serving couples requests through shared GPU execution, one adversarial input can also amplify tail latency and reduce effective throughput for co-scheduled benign users, echoing the broader resource-exhaustion pattern seen in sponge attacks and recent LLM availability attacks~\citep{shumailov2021sponge,zhang2024safeguard,dong2024engorgio,wang2026latencydos}. 

%% file: Sections/Methodology.tex
\section{Attack Method}
\label{sec:method}

\begin{figure}[t]
\centering
\includegraphics[width=\linewidth]{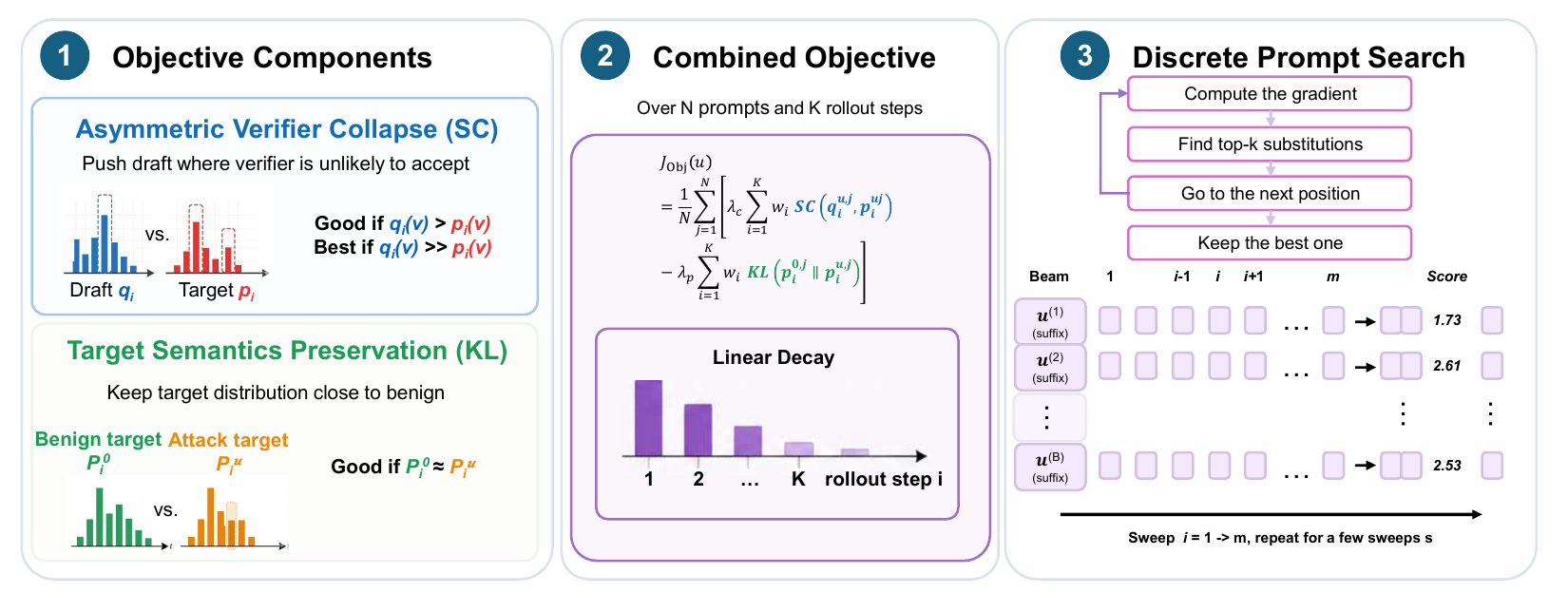}
\caption{Overview of the ADSD pipeline. ADSD optimizes a short discrete
suffix over a calibration batch by combining a speculative-decoding-aligned
Soft-Collapse objective with a target preservation objective, and solves the
resulting search problem using gradient-guided sequential beam search.}
\label{fig:adsd-pipeline}
\end{figure}

We propose ADSD, as shown in Figure~\ref{fig:adsd-pipeline}, which targets the internal acceptance mechanism of speculative decoding while
preserving the final target-model response. At rollout step \(i\), let
\(q_i(\cdot)\) and \(p_i(\cdot)\) denote the draft and target next-token
distributions under the same decoding context. For a drafted token \(v\), the
verifier accepts it with probability
\begin{equation}
    \alpha_i(v)
    =
    \min\!\left(1,\frac{p_i(v)}{q_i(v)}\right).
    \label{eq:acceptance-rule}
\end{equation}

This rule is asymmetric. If \(q_i(v)\le p_i(v)\), the token is accepted with
probability one; increasing disagreement in this direction does not reduce
acceptance. Rejection pressure only appears when the draft is overconfident,
i.e., when \(q_i(v)>p_i(v)\). Writing the draft-target log margin as
\[
    \Delta_i(v)=\log q_i(v)-\log p_i(v),
\]
the same verifier rule becomes
\begin{equation}
    \alpha_i(v)
    =
    \exp\!\left(-\mathrm{ReLU}(\Delta_i(v))\right).
    \label{eq:log-acceptance}
\end{equation}

Thus, the useful attack signal is the positive log-margin: tokens with
\(\Delta_i(v)>0\) are those that the verifier is less likely to accept.

\paragraph{Soft-Collapse.}
Motivated by this, we propose the Soft-Collapse objective:
\begin{equation}
    \mathrm{SC}(q_i,p_i)
    =
    \sum_{v}
    q_i(v)\,
    \mathrm{ReLU}
    \bigl(
        \log q_i(v)-\log p_i(v)
    \bigr).
    \label{eq:soft-collapse}
\end{equation}
Soft-Collapse keeps only the verifier-relevant side of draft-target disagreement.
This makes SC a compact verifier-aligned margin objective for collapsing speculative acceptance.

\paragraph{Stealth via Target Semantics Preservation.}

A suffix that only maximizes collapse may also corrupt the target model's own
continuation distribution, making the attack visible through degraded answers.
ADSD therefore anchors the attacked target distribution to the benign target
distribution. Let \(p_i^{0}\) be the target distribution without the adversarial suffix and \(p_i^{u}\) be the target distribution under a candidate suffix \(u\). We preserve
target behavior with the forward KL penalty
\begin{equation}
    \mathrm{KL}(p_i^{0}\Vert p_i^{u})
    =
    \sum_{v}
    p_i^{0}(v)
    \bigl(
        \log p_i^{0}(v)-\log p_i^{u}(v)
    \bigr).
    \label{eq:target-preserve-kl}
\end{equation}
This term discourages the suffix from suppressing tokens that the benign target
model would likely generate, preventing the attack from trading efficiency
collapse for obvious semantic corruption.

\paragraph{Final Objective.}

ADSD optimizes a universal suffix \(u\) over \(N\) calibration prompts and \(K\)
rollout steps. For prompt \(x^{(j)}\), let \(q_i^{u,j}\) and \(p_i^{u,j}\) be the
draft and target distributions under suffix \(u\), and let \(p_i^{0,j}\) be the
benign target distribution. The complete objective is
\begin{equation}
J_{\mathrm{ADSD}}(u)
=
\frac{1}{N}
\sum_{j=1}^{N}
\sum_{i=1}^{K}
w_i
\left[
    \lambda_c\,
    \mathrm{SC}
    \left(
        q_i^{u,j},
        p_i^{u,j}
    \right)
    -
    \lambda_p\,
    \mathrm{KL}
    \left(
        p_i^{0,j}
        \Vert
        p_i^{u,j}
    \right)
\right].
\label{eq:adsd-objective}
\end{equation}
Here, \(w_i\propto K-i+1\) gives larger weight to earlier rollout
positions. Early rejection is especially costly because it prevents the system
from benefiting from the remaining draft tokens in the same speculative block. 
The SC term pushes draft mass into
the verifier's rejection-relevant region, while the KL term keeps the target model
close to its benign behavior. ADSD therefore attacks the acceleration path rather
than simply damaging the final answer.

\paragraph{Search over Discrete Suffixes.}

The suffix \(u=(u_1,\ldots,u_m)\) is discrete, so ADSD uses gradient-guided
sequential beam search. During sweep \(s\) and coordinate \(r\), we maintain a beam
of candidate suffixes. For each candidate \(u\), we compute the gradient of
\(J_{\mathrm{ADSD}}\) with respect to the embedding of the current suffix token:
\[
    g_r(u)
    =
    \nabla_{e(u_r)}
    J_{\mathrm{ADSD}}(u).
\]
ADSD keeps the top-ranked replacement tokens, evaluates the full objective after each discrete substitution,
and retains the best \(B\) suffixes in the beam. We sweep left-to-right over all
\(m\) suffix positions and repeat for a small number of sweeps. Finally, we pick the one with the highest objective value.

%% file: Sections/Eval.tex
 \section{Evaluation}
\label{sec:eval}

\subsection{Experimental Setup}
\noindent\textbf{Implementation Details.} We perform our evaluation on the  GSM8K~\citep{cobbe2021trainingverifierssolvemath} dataset, with a white-box target-draft pair using the widely adopted GPTQ-quantized 8-bit instruction-tuned Qwen2.5 series~\citep{bai2023qwen}. The main Qwen experiments are run on a single A100 80GB GPU. The LLaMA-3 70B $\rightarrow$ 8B evaluation uses a tensor-parallel Palmetto testbed with four nodes and two A100 80GB GPUs per node. The EAGLE-3 experiments are run separately on a local single RTX 4090 GPU. The Qwen and LLaMA GSM8K evaluations use the same HSD few-shot prompt followed by the model chat template, while EAGLE-3~\citep{li2025eagle3} follows native turns and a system prompt. To demonstrate generalizability, we evaluate HumanEval~\citep{chen2021evaluatinglargelanguagemodels} and CNN/DailyMail~\citep{see-etal-2017-get} using adversarial suffixes calibrated exclusively on GSM8K. We also scale the target models to other architectures, including Qwen-32B and LLaMA-3-70B, to validate cross-architecture vulnerability.

\noindent\textbf{Baseline Methods.} We compare our method against three baselines: (1) Benign input, (2) randomly generated discrete suffix, (3) a state-of-the-art gradient-guided discrete search (adapted GCG)~\citep{zou2023universal}. The impact on systems-level performance is quantified using mean sample time (average wall-clock time required to generate one complete answer), block efficiency (the average number of target-accepted tokens per speculative verification block), decoding speed, and task quality metrics.

\subsection{Baseline Comparison}

Once the main GSM8K vulnerability is established, the critical question is whether the specialized ADSD suffix-search framework is actually useful. Table~\ref{tab:baseline-comparison} compares ADSD against two attack baselines and the benign speculative-decoding baseline on the same tokenwise GSM8K benchmark. 

\begin{table}[h]
\centering
\caption{Tokenwise GSM8K results for benign decoding and three prompt-search attacks. The key comparison is whether a searched suffix causes a large latency increase while keeping task accuracy close to the benign baseline.}
\label{tab:baseline-comparison}
\footnotesize
\begin{tabular}{lccccc}
\toprule
Method & Time (s) & Block eff. & Speed (tok/s) & GSM8K acc. \\
\midrule
Benign baseline & 26.05 & 6.028 & 81.19 & 0.821 \\
Random search & 30.23 (+16.0\%) & 5.021 (-16.7\%) & 71.53 (-11.9\%) & 0.701 (-14.6\%)\\
GCG-adapted~\cite{zou2023universal} & 35.83 (+37.5\%) & 4.578 (-24.1\%) & 64.13 (-21.0\%) & 0.821 \\
\rowcolor{lightblue}
\textbf{ADSD} & \textbf{42.29 (+62.3\%)} & \textbf{3.683 (-38.9\%)} & \textbf{51.91 (-36.1\%)} & 0.802 \\
\bottomrule
\end{tabular}
\end{table}

This comparison shows the key advantage of our approach. While both baselines find suffixes that perturb calibration-time verifier behavior, only ADSD produces a strong end-to-end slowdown on the actual benchmark. Random search reaches 30.23s and the adapted GCG baseline reaches 35.83s, both well below ADSD's 42.29s. ADSD therefore produces the largest sustained degradation of the speculative mechanism, reducing block efficiency from 6.028 to 3.683 and decoding speed from 81.19 tok/s to 51.91 tok/s, while incurring only a modest GSM8K accuracy drop.

\subsection{Attack Under Different Speculative Design Strategies}

Modern inference engines deploy more advanced speculative mechanisms than simple token-by-token verification. Table~\ref{tab:spec-designs} evaluates whether ADSD's prompt-side attack can bypass these architectural improvements by targeting Blockwise decoding and Hierarchical Speculative Decoding (HSD)~\citep{zhou2026hsd}.

\begin{table}[h]
\centering
\caption{ADSD under three speculative decoding designs. Each attack row reports the change relative to the benign row above it and uncertainty ($\pm$) over the GSM8K questions for attack results.}
\label{tab:spec-designs}
\footnotesize
\begin{tabular}{lccccc}
\toprule
Method & Setting & Time (s) & Block eff. & Speed (tok/s) & GSM8K acc. \\
\midrule
Tokenwise & Benign & 26.05 & 6.028 & 81.19 & 0.821 \\
\rowcolor{lightblue}
Tokenwise & \textbf{ADSD} & \textbf{42.29$\pm$2.18 (+62.3\%)} & \textbf{3.683$\pm$0.118 (-38.9\%)} & \textbf{51.91$\pm$1.54 (-36.1\%)} & 0.802 \\
Blockwise & Benign & 25.39 & 6.048 & 83.97 & 0.825 \\
\rowcolor{lightblue}
Blockwise & \textbf{ADSD} & \textbf{41.83$\pm$2.41 (+64.8\%)} & \textbf{3.667$\pm$0.111 (-39.4\%)} & \textbf{51.57$\pm$1.42 (-38.6\%)} & 0.825 \\
HSD~\citep{zhou2026hsd} & Benign & 24.76 & 6.302 & 85.26 & 0.833 \\
\rowcolor{lightblue}
HSD~\citep{zhou2026hsd} & \textbf{ADSD} & \textbf{41.12$\pm$2.35 (+66.1\%)} & \textbf{3.836$\pm$0.127 (-39.1\%)} & \textbf{53.93$\pm$1.78 (-36.7\%)} & 0.833 \\
\bottomrule
\end{tabular}
\end{table}

Advanced strategies like HSD~\citep{zhou2026hsd} achieve slightly better benign block efficiency (6.302), but they remain highly vulnerable to the exact same adversarial suffix. Under ADSD, Blockwise latency rises from 25.39s to 41.83s (+64.8\%), and HSD~\citep{zhou2026hsd} latency rises from 24.76s to 41.12s (+66.1\%). In both cases, block efficiency collapses to roughly the same range as the tokenwise attack, suggesting that these design changes do not remove the draft-target divergence exploited by ADSD.

\subsection{Transferability to New Domains}

We next test whether a suffix optimized only on GSM8K transfers to different tasks without task-specific re-optimization. As shown in Table~\ref{tab:transfer-benchmark}, the suffix transfers strongly to structured code generation. On HumanEval, it increases mean sample time by 141.8\%, while also reducing syntax validity from 0.970 to 0.823 and pass@1 from 0.829 to 0.683. The same suffix also transfers to long-form abstractive summarization on CNN/DailyMail, producing a 30.4\% slowdown.

\begin{table}[h]
\centering
\caption{Cross-domain transfer of GSM8K-generated ADSD adversarial suffix. The table shows attack results without any task-specific re-optimization.}
\label{tab:transfer-benchmark}
\footnotesize
\begin{tabularx}{\linewidth}{lccc>{\raggedright\arraybackslash}X}
\toprule
\multicolumn{2}{l}{HumanEval~\citep{chen2021evaluatinglargelanguagemodels}}\\
\midrule
Setting & Time (s) & Block eff. & Speed (tok/s) & Task quality \\
\midrule
Benign & 13.14  & 5.890 & 80.50 & Syn. valid = 0.970, pass@1 = 0.829 \\
\rowcolor{lightblue}
\textbf{ADSD} & \textbf{31.77 (+141.8\%)} & \textbf{3.968 (-32.6\%)} & \textbf{55.94 (-30.5\%)} & Syn. valid = 0.823, pass@1 = 0.683 \\
\midrule
\multicolumn{2}{l}{CNN/DailyMail~\citep{see-etal-2017-get}}\\
\midrule
Setting & Time (s) & Block eff. & Speed (tok/s) & Task quality \\
\midrule
Benign & 34.68 & 2.538 & 37.72 & R-1 = 0.227, R-L = 0.204 \\
\rowcolor{lightblue}
\textbf{ADSD} & \textbf{45.24 (+30.4\%)} & \textbf{2.027 (-20.1\%)} & \textbf{30.57 (-18.9\%)} & R-1 = 0.048, R-L = 0.043 \\
\bottomrule
\end{tabularx}
\vspace{-15pt}
\end{table}

We attribute the variance in attack severity across these domains to the complexity difference between writing code and writing summaries. Code generation (HumanEval) is highly rigid, relying on exact syntax, fixed keywords, and strict formatting. Speculative decoding excels here under normal conditions because the draft model can easily guess predictable code blocks. Our attack disrupts this precise alignment, forcing the draft to propose more false overconfident tokens and causing the verifier to reject much more aggressively, which leads to large latency spikes. Conversely, long-form summarization (CNN/DailyMail) is naturally open-ended. Because there are many valid phrasings, the draft and target models already disagree more often on exact word choice, resulting in a much lower benign speedup (2.538 block efficiency). The attack therefore has less headroom to further degrade performance, but it still consistently reduces efficiency. However, the same GSM8K-optimized suffix substantially harms task quality on both transfer datasets, indicating that cross-domain latency transfer does not necessarily imply cross-domain stealth.

\subsection{Cross-LLM Architecture Evaluation}

We next evaluate whether ADSD-induced acceptance collapse is specific to the
main Qwen2.5 14B $\rightarrow$ 0.5B setting, or whether similar behavior appears
under other draft-target pairs and speculative decoding architectures.
Table~\ref{tab:scaling-placeholder} reports results for two Qwen2.5 target
sizes using the same 0.5B draft model, while Table~\ref{tab:combined-llama-eagle}
extends the evaluation to a LLaMA-3 draft-target pair and to EAGLE-3.

\begin{table}[h]
\vspace{-10pt}  
\centering
\caption{Cross-architecture transfer on larger target models. Attack rows include the change relative to the benign baseline for the same draft-target pair, making the slowdown and efficiency drop easy to compare.}
\label{tab:scaling-placeholder}
\footnotesize
\begin{tabular}{lccccc}
\toprule
Target $\rightarrow$ Draft & Setting & Time (s) & Block eff. & Speed (tok/s) & GSM8K acc. \\
\midrule
Qwen2.5 14B $\rightarrow$ 0.5B & Benign & 26.05 & 6.028 & 81.19 & 0.821 \\
\rowcolor{lightblue}
Qwen2.5 14B $\rightarrow$ 0.5B & \textbf{ADSD} & \textbf{42.29 (+62.3\%)} & \textbf{3.683 (-38.9\%)} & \textbf{51.91 (-36.1\%)} & 0.802 \\
Qwen2.5 32B $\rightarrow$ 0.5B & Benign & 39.26 & 6.139 & 53.21 & 0.829 \\
\rowcolor{lightblue}
Qwen2.5 32B $\rightarrow$ 0.5B & \textbf{ADSD} & \textbf{52.84 (+34.6\%)} & \textbf{4.105 (-33.1\%)} & \textbf{41.32 (-22.3\%)} & 0.821 \\

\bottomrule
\end{tabular}
\end{table}
\begin{wraptable}{r}{0.6\textwidth}
  \centering
  \footnotesize
  
  \caption{ADSD attack on LLaMA-3 (70B $\rightarrow$ 8B) and EAGLE-3~\citep{li2025eagle3}. The attack substantially degrades acceleration in both architectures.}
  \vspace{5 pt}
  \label{tab:combined-llama-eagle}
  \begin{tabular}{lccc}
    \toprule
    Setting & Blk. eff. & Speed (tok/s) & GSM8K \\
    \midrule
    \multicolumn{4}{l}{\textbf{LLaMA (70B $\rightarrow$ 8B)}}\\
    \midrule
    Benign               & 6.145 & 29.33  & 0.872 \\
    \rowcolor{lightblue}
    \textbf{ADSD} & \textbf{4.509 ($-$26.6\%)} & \textbf{21.40 ($-$27.0\%)} & 0.802 \\
    \midrule
    \multicolumn{4}{l}{\textbf{EAGLE-3}~\citep{li2025eagle3}}\\
    \midrule
    Benign                & 3.926 & 116.81 & 0.638 \\
    \rowcolor{lightblue}
    \textbf{ADSD} & \textbf{0.908 ($-$76.9\%)} & \textbf{26.95 ($-$76.9\%)} & 0.575 \\
    \bottomrule
  \end{tabular}
\end{wraptable}

For Qwen2.5 32B $\rightarrow$ 0.5B, the same attack pattern persists. The relative latency increase is smaller than in the 14B setting, but the direction of the effect is consistent. The LLaMA-3 and EAGLE-3 results further suggest that prompt-induced acceptance collapse can occur beyond independent Qwen2.5 draft-target pairs. In the
LLaMA-3 70B $\rightarrow$ 8B setting, ADSD reduces block efficiency by 26.6\%
and decoding speed by 27.0\%. In EAGLE-3, which uses a different speculative
architecture, the degradation is larger: both block efficiency and decoding
speed fall by 76.9\%.

\begin{table}[t]
\centering
\vspace{-5pt}  
\caption{Ablation of ADSD objective terms. The table shows which component is responsible for the computational slowdown and which component keeps the final output useful.}
\label{tab:ablation-placeholder}
\footnotesize
\begin{tabular}{lcccc}
\toprule
Objective variant & Time (s) & Block eff. & Speed (tok/s) & Accuracy \\
\midrule
\rowcolor{lightblue}
\textbf{Full ADSD objective} & \textbf{42.29} & \textbf{3.683} & \textbf{51.91} & \textbf{0.802} \\
w/o soft-collapse & 27.66 (-39.4\%) & 5.889 (+60.5\%) & 79.97 (+58.0\%) & 0.821  \\
w/o target-preserve KL & 49.67 (+8.9\%) & 3.459 (-5.7\%) & 49.28 (-2.7\%) & 0.703 (-12.3\%) \\
\bottomrule
\end{tabular}
\vspace{-10pt}  
\end{table}

\subsection{Ablation Study}

Table~\ref{tab:ablation-placeholder} ablates the primary terms of our formulation: the Soft-Collapse surrogate and the target-preserving KL divergence. When the Soft-Collapse component is removed, the attack fails entirely; execution time falls back to 27.66s (nearly matching the benign baseline), and block efficiency recovers to 5.889. This demonstrates that standard continuous optimization lacks the directional guidance required to systematically push draft tokens into the verifier's rejection-relevant positive-margin region. Conversely, removing the target-preserve KL term results in a highly effective computational attack (49.67s), but heavily corrupts the output, causing task accuracy to plummet to 0.703. This violates the stealth constraints of the threat model, suggesting that both terms are important for achieving a large slowdown while
avoiding severe GSM8K accuracy degradation in this setting.

%% file: Sections/Discussion.tex
\section{Defenses and Limitations}
\label{sec:defenses}
\vspace{-5pt}
The closest existing defense family to our threat model is prompt-side anomaly filtering, including perplexity-based detection and related input-screening heuristics~\citep{alon2023perplexity,jain2023baseline,zou2023universal}. More broadly, recent work on jailbreak defense has repeatedly found a tension between safety gains and over-defensiveness or deployment overhead~\citep{phute2023selfdefense,varshney2024art}. Our results suggest that this defense family is only a partial fit for speculative decoding. ADSD uses a very short injected suffix, but it is evaluated inside a much longer benign prompt, often with few-shot context. This makes \emph{global} prompt perplexity a weak signal, because the anomalous suffix is diluted by the surrounding text. More localized checks may recover some sensitivity, but they also add extra computational overhead and cost and may inject false-positive warnings. 

Output-based checks are also insufficient on their own. Safety-aware decoding and self-reflection style defenses primarily intervene on the generated answer or on the model's own judgment of that answer~\citep{xu2024safedecoding,phute2023selfdefense}, but ADSD targets a different failure mode. As shown in Appendix~\ref{app:case-study-draft-vs-spec}, the same adversarial suffix that drives the standalone draft model into obviously irrelevant text can still leave the full speculative system with a semantically acceptable final answer after a short corrupted suffix. This means user-visible answer quality and internal verifier cost can decouple: the system may still look correct while speculative efficiency has already collapsed. That observation is important for defense, because it implies that post-hoc answer filtering, semantic quality checks, or simple output screening can miss the attack precisely when the target model repairs the draft's visible mistakes. Our analysis is still limited to white-box optimization and a representative set of draft-target pairs, but the central lesson is robust: checking only the input or only the final output is unlikely to fully address prompt-induced verifier collapse, because the attack exploits the draft-target interaction inside speculative decoding rather than only the surface form of the prompt or the answer.

We leave defense design and evaluation to future work. One practical mitigation direction is online acceptance-rate monitoring: if the accepted-token rate drops sharply relative to a prompt-class baseline, the server can disable speculation or route the request to target-only decoding. This does not prevent the first attacked request from incurring additional overhead, but it can bound repeated cost amplification.

%% file: Sections/Conclusion.tex
\section{Conclusion}
\vspace{-5pt}

We introduced ADSD, a prompt-side attack that exposes a previously underexplored algorithmic vulnerability in speculative decoding by systematically enforcing draft rejection. By utilizing the Soft-Collapse surrogate objective to navigate the verifier's asymmetric acceptance rule, ADSD effectively collapses verifier efficiency. Empirically, our optimized adversarial suffix increases mean sample time on GSM8K from 26.05s to 42.29s while largely preserving GSM8K accuracy. Furthermore, we evaluate this attack design on different datasets and architectures, and show that the attack is consistently effective. These results demonstrate that lossless output semantics do not guarantee operational cost robustness; by transforming speculative acceleration into a prompt-controlled attack surface, ADSD shows that prompt-induced verifier collapse is a recurring vulnerability across the evaluated speculative-decoding settings, and this highlights the need for further exploration.

%% file: Sections/Appendix.tex
\newpage

\appendix

\section{Additional Evaluation Settings}
\label{sec:appendix-gsm8k-benchmark}
\paragraph{Reproducibility details.}
Unless otherwise noted, ADSD uses the exact checkpoints recorded in the run metadata: Qwen2.5-14B-Instruct-GPTQ-Int8 as target with Qwen2.5-0.5B-Instruct-GPTQ-Int8 as draft for the main Qwen results; Qwen2.5-32B-Instruct with the same 0.5B draft for the larger-Qwen transfer; Llama-3.1-70B-Instruct as target with Llama-3.1-8B-Instruct as draft for the LLaMA experiments; and EAGLE3-LLaMA3.1-Instruct-8B for the EAGLE-3 experiments. Qwen and LLaMA GSM8K evaluations use the few-shot wrapper followed by the model chat template; EAGLE-3 follows its native pipeline with a system prompt. For standard tokenwise speculative decoding, the speculative block size is $\gamma=10$, temperature is $1.0$, \texttt{top\_k}=0, generation uses sampling as required by the underlying implementation, and \texttt{max\_new\_tokens}=512. For LLaMA tensor-parallel evaluation, we use the same $\gamma=10$, temperature $1.0$, and \texttt{max\_new\_tokens}=512. For EAGLE-3 evaluation, we use GSM8K dataset and temperature $1.0$ under the native EAGLE-3 evaluator. Main Qwen experiments are run on one A100 80GB GPU; LLaMA-3 70B$\rightarrow$8B uses four nodes with two A100 80GB GPUs per node; and the local EAGLE-3 runs are executed on a single RTX 4090 GPU. 

\paragraph{ADSD search hyperparameters.}
For the suffix search, we optimize a suffix of length $m=5$ over a calibration set of $N=10$ GSM8K training prompts, selected with calibration offset $1$ and stride $100$. Each calibration prompt uses a target rollout horizon of $K=64$ tokens. The discrete search uses a beam size of $4$, per-position candidate set size $8$, and $5$ left-to-right sweeps. The rollout weighting uses normalized linear decay over positions. In the main revised Qwen and EAGLE runs, the objective keeps only the Soft-Collapse term and the target-preservation term: collapse weight $\lambda_c=2.0$, target-preserve weight $\lambda_p=1.0$ for Qwen and $\lambda_p=2.0$ for the EAGLE-3 fix-align run. Initialization uses random initialization. Evaluation outputs are not length-matched across methods; instead, all methods are run with the same generation budget and task-specific quality is measured on the resulting completions.

\section{Additional Attack Analysis}
\paragraph{Example attacked input prompt.}
For completeness, we show the exact Qwen-style input prompt format as an example. The adversarial suffix is not expanded here; instead, we mark its insertion point with a placeholder. The same final user question and suffix placement are used for both the standalone draft run and the full speculative run.

\begin{quote}
\small\ttfamily
<|im\_start|>system\\
You are Qwen, created by Alibaba Cloud. You are a helpful assistant.<|im\_end|>\\
<|im\_start|>user\\
Question: There are 15 trees in the grove. Grove workers will plant trees in the grove today. After they are done, there will be 21 trees. How many trees did the grove workers plant today?\\
Let's think step by step\\
There are 15 trees originally.\\
Then there were 21 trees after some more were planted.\\
So there must have been 21 - 15 = 6.\\
The answer is 6.\\
\\
Question: If there are 3 cars in the parking lot and 2 more cars arrive, how many cars are in the parking lot?\\
Let's think step by step\\
There are originally 3 cars.\\
2 more cars arrive.\\
3 + 2 = 5.\\
The answer is 5.\\
\\
Question: Leah had 32 chocolates and her sister had 42. If they ate 35, how many pieces do they have left in total?\\
Let's think step by step\\
Originally, Leah had 32 chocolates.\\
Her sister had 42.\\
So in total they had 32 + 42 = 74.\\
After eating 35, they had 74 - 35 = 39.\\
The answer is 39.\\
\\
Question: Jason had 20 lollipops. He gave Denny some lollipops. Now Jason has 12 lollipops. How many lollipops did Jason give to Denny?\\
Let's think step by step\\
Jason started with 20 lollipops.\\
Then he had 12 after giving some to Denny.\\
So he gave Denny 20 - 12 = 8.\\
The answer is 8.\\
\\
Question: Shawn has five toys. For Christmas, he got two toys each from his mom and dad. How many toys does he have now?\\
Let's think step by step\\
Shawn started with 5 toys.\\
If he got 2 toys each from his mom and dad, then that is 4 more toys.\\
5 + 4 = 9.\\
The answer is 9.\\
\\
Question: There were nine computers in the server room. Five more computers were installed each day, from monday to thursday. How many computers are now in the server room?\\
Let's think step by step\\
There were originally 9 computers.\\
For each of 4 days, 5 more computers were added.\\
So 5 * 4 = 20 computers were added.\\
9 + 20 is 29.\\
The answer is 29.\\
\\
Question: Michael had 58 golf balls. On tuesday, he lost 23 golf balls. On wednesday, he lost 2 more. How many golf balls did he have at the end of wednesday?\\
Let's think step by step\\
Michael started with 58 golf balls.\\
After losing 23 on tuesday, he had 58 - 23 = 35.\\
After losing 2 more, he had 35 - 2 = 33 golf balls.\\
The answer is 33.\\
\\
Question: Olivia has \$23. She bought five bagels for \$3 each. How much money does she have left?\\
Let's think step by step\\
Olivia had 23 dollars.\\
5 bagels for 3 dollars each will be 5 x 3 = 15 dollars.\\
So she has 23 - 15 dollars left.\\
23 - 15 is 8.\\
The answer is 8.\\
\\
Question: Janet's ducks lay 16 eggs per day. She eats three for breakfast every morning and bakes muffins for her friends every day with four. She sells the remainder at the farmers' market daily for \$2 per fresh duck egg. How much in dollars does she make every day at the farmers' market?\\
<|im\_end|>\\
<|im\_start|>assistant
[ADV\_SUFFIX]\\
\end{quote}

\subsection{Case Study: Standalone Draft Failure vs. Speculative Recovery}
\label{app:case-study-draft-vs-spec}

We next show a concrete example of why ADSD can be computationally harmful while still preserving many final answers. For each case below, we use the exact same few-shot GSM8K prompt template, the same user question, and the same injected adversarial suffix. The only difference is whether we run (i) the standalone draft model alone or (ii) the full speculative decoding system with a larger target verifier. To avoid distracting the reader, we display only the generated continuation and omit any exact echoed copy of the injected suffix at the start of the answer.

\paragraph{Example 1: Qwen2.5-14B target with Qwen2.5-0.5B draft.}
Under the attacked prompt, the standalone 0.5B draft model is immediately derailed and generates irrelevant code-related text:
\begin{quote}
\small\ttfamily
The correct way is to write:\\
backgroundResource="@drawable/box"\\
...\\
In Android Studio, you can choose the background resource from the property panel ...
\end{quote}

In contrast, the speculative system also exhibits a visibly corrupted prefix, but then recovers and reaches the correct answer:
\begin{quote}
\small\ttfamily
(stepBgResId)\\
Janet's ducks lay 16 eggs per day.\\
She eats 3 eggs every day for breakfast.\\
She also uses 4 eggs every day to bake muffins.\\
So the number of eggs she uses is 3 + 4 = 7.\\
This means she sells 16 - 7 = 9 eggs.\\
Since each egg sells for \$2, she makes 9 * \$2 = \$18.\\
The answer is \$18.
\end{quote}

\paragraph{Example 2: Qwen2.5-32B target with Qwen2.5-0.5B draft.}
The standalone 0.5B draft model again fails immediately and emits a malformed code fragment:
\begin{quote}
\small\ttfamily
fromstring(res, "decimal")];
\end{quote}

The speculative system still shows an anomalous prefix, but the larger target model recovers the reasoning path:
\begin{quote}
\small\ttfamily
convert and calculate step by step:\\
Janet's ducks lay 16 eggs per day.\\
She eats 3 eggs herself.\\
She uses another 4 eggs for baking.\\
So she sells 16 - 3 - 4 = 9 eggs.\\
At \$2 per egg, she makes \$18.\\
The answer is \$18.
\end{quote}

\paragraph{Takeaway.}
These examples highlight an important asymmetry. The adversarial suffix can strongly derail the standalone draft model, often causing it to generate irrelevant or nonsensical text. Yet in the full speculative pipeline, the larger target model may still repair the corrupted trajectory and recover a correct final answer. This makes ADSD operationally dangerous: the attack can collapse speculative efficiency even when the final user-visible answer remains correct enough to avoid simple quality-based detection.